\documentclass[twocolumn,showpacs,pre,floatfix,eps]{revtex4}
\ifx\pdfoutput\undefined
\usepackage{graphicx}
\else
\usepackage{epstopdf}
\usepackage{epsfig}
\usepackage{color}

\usepackage{amssymb,amsmath,stmaryrd,tabularx}
\usepackage{wrapfig}
\usepackage{bm,natbib}
\fi
\usepackage[center]{subfigure}

\newcommand{\be}{\begin{equation}}
\newcommand{\ee}{\end{equation}}
\newcommand{\bea}{\begin{eqnarray}}
\newcommand{\eea}{\end{eqnarray}}

\begin{document}
\title{The $Re$-number dependence of the longitudinal dispersion in a turbulent channel flow}
\author{Christopher Hawkins$^1$, Luiza Angheluta$^1$, Marcin Krotkiewski$^2$, and Bj{\o}rn Jamtveit$^2$}
\affiliation{
$^1$Physics of Geological Processes, Department of Physics, \\ University of Oslo, P.O. 1048 Blindern, 0316 Oslo, Norway\\
$^2$Physics of Geological Processes, Department of Geoscience, \\ University of Oslo, P.O. 1048 Blindern, 0316 Oslo, Norway}

\pacs{47.27.tb, 47.27.nd, 47.11.-j, 47.27.E-}

\begin{abstract}
In Taylor's theory, the longitudinal dispersion in turbulent pipe flows approaches, on long timescales, a diffusive behavior with a constant diffusivity $K_L$, that depends \emph{empirically} on the Reynolds number $Re$. We show that the dependence on $Re$ can be determined from the turbulent energy spectrum. By using the intimate connection between the friction factor and longitudinal dispersion in wall-bounded turbulence, we predict different asymptotic scaling laws of $K_L(Re)$ depending on the different turbulent cascades in two-dimensional turbulence. We also explore numerically the $K_L(Re)$ dependence in turbulent channel flows with smooth and rough walls using a lattice Boltzmann method. 
\end{abstract}
\maketitle
\date{}

\section{Introduction}
Wall-bounded turbulent flows enhance longitudinal dispersion of matter due to the combined effect of velocity fluctuations and mean shear. In two seminal papers on longitudinal dispersion of passive matter in laminar and turbulent pipe flows~\cite{Taylor1953,Taylor1954}, Taylor predicted that, on long timescales, longitudinal spreading of matter in a straight pipe can be described by a one dimensional diffusive process with an effective diffusivity coefficient that is many orders of magnitude larger than the molecular one. In contrast to the pair dispersion that is enhanced mostly by turbulent fluctuations such that on inertial scales it behaves superdiffusive~\cite{salazar2009two}, the longitudinal (single-particle) dispersion is strongly influenced by the mean flow properties. Therefore, wall shear and boundary layers play a key role in the transport of matter. Despite numerous studies on passive advection~\cite{kraichnan1968small,chertkov1995normal,bernard1996anomalous,warhaft2000passive}, the fact remains that there is a lack of fundamental understanding of the dependence of the longitudinal dispersion coefficient $K_L$ on the Reynolds number $Re$, beyond empirical evidence~\cite{tichacek1957axial,flint1969longitudinal,ekambara2003axial}. Enhanced longitudinal dispersion is a ubiquitous natural phenomenon and has an immediate impact on estimating flow rates and mixing in long pipelines~\cite{tichacek1957axial,fischer1973longitudinal,meier2014microstructural}, as well on transport and deposition/sedimentation conditions in natural flows, e.g.~\cite{lacasce2008statistics,wilson1996review,fischer1973longitudinal}. 

In this paper, we aim to provide a more fundamental understanding of the observed scaling law of $K_L$ with $Re$-number, by relating it to the inertial scaling law of the turbulence energy spectrum. Within Taylor's theory, the longitudinal dispersion coefficient $K_L$ is directly related to the wall frictional shear stress, which is related to the friction factor, $f$~\cite{Taylor1954}. This is a remarkable connection between a measure of bulk transport of matter and a measure of flow resistance by a shear stress exerted on a wall. It implies that when Taylor's theory of dispersion applies, the transport properties of momentum and matter are related to each other beyond Reynolds analogy. As a consequence of this connection, we show that the spectral link of the friction factor to the turbulent energy spectrum originally proposed in Ref.~\cite{gioia2006turbulent} can be extended to scalar dispersion. It means that a turbulent state characterized by a given turbulent spectrum determines not only the properties of the wall friction, but also the dispersion of matter in the pipe. The other way around, by accessing the dispersion and wall friction properties, we can infer about the turbulent state.   

The rest of the paper is organized as follows. In Section II, using Taylor's approach we calculate the longitudinal dispersion coefficient in a turbulent channel flow. Its connection to the energy spectrum is discussed in Section III, followed by a description of the numerical approach using Lattice Boltzmann in Section IV. The results are discussed in the final Section V.  

\section{Dispersion in turbulent channel flow}
Following Taylor's approach~\cite{Taylor1954}, we formulate the dispersion in two-dimensional wall-bounded turbulence, and include the boundary layer effect on the mean velocity profile, hence on the dispersion law with $Re$ number. Albeit, they maybe confined in thin regions near the walls, the boundary layers tend to concentrate more tracer particles because of the reduced mean velocity, and therefore alter the global dispersion~\cite{tichacek1957axial,flint1969longitudinal,ekambara2003axial}. 

We start with the scalar advection equation for the concentration field $c(x,y,t)$ of dispersed tracers 
\begin{equation}\label{eq:1}
\frac{\partial c}{\partial t}+\mathbf{u}\cdot\nabla c = 0, 
\end{equation}
where $\mathbf{u}(x,y,t)$ denotes the incompressible turbulent fluid velocity, and molecular diffusion is neglected compared to the advective transport.

On timescales larger than the integral scale, small-scale turbulent fluctuations become statistically uncorrelated and can be separated from the mean flow by Reynolds decomposition.~For a statistically stationary, but anisotropic and inhomogeneous flow, the Reynolds decomposition is applied in the comoving frame relative to the mean flow direction, $\xi = x-U_0t$, where $U_0$ is the mean velocity obtained by space and time averaging of $\mathbf{u}(x,y,t)$, hence $\mathbf{u} = \overline{\mathbf{U}}(y)+\mathbf{u'}(\xi,y,t)$, where the time-averaged velocity $\overline{\mathbf{U}}(y) = U_0+U_x(y)\mathbf{e}_x$ has the global average velocity $U_0$ part and the steady-state mean velocity profile in the comoving frame $U_x(y)$. By analogy, the particle concentration is split into a time-averaged part and fluctuations $c = \overline{C}(y)+c'(\xi,y,t)$. Upon substitution and time-averaging, the advection equation becomes 
\begin{equation}\label{eq:c2}
U_x(y)\frac{\partial \overline{C}}{\partial\xi}+\nabla\cdot(\overline{\mathbf{u'}c'}) = 0. 
\end{equation}
To proceed further with Eq.~(\ref{eq:c2}), a closure assumption for the turbulent flux $\overline{\mathbf{u'}c'}$ is need. In Taylor's theory of dispersion, the \emph{Reynolds analogy} between transport of momentum and matter is used as a first assumption. For wall-bounded turbulent flows, this assumption is valid in the~\lq outer\rq~layer, i.e. outside the boundary layers where the flow is well-mixed so that the velocity and concentration profiles become universal. Then the turbulent shear stress $\tau(y)$ and the Reynolds flux terms ($\overline{u'_xu'_y} $ and $\overline{\mathbf{u'}c'}$) are large compared with molecular diffusion through mean gradients, and follow a Fickian law as  
\begin{eqnarray}\label{eq:Reynolds}
\overline{\mathbf{u'} c'} &=& -\epsilon(y)\nabla \overline{C}\\
\frac{1}{\rho}\tau(y) &=& -\epsilon(y) U_x'(y),
\end{eqnarray}
where the local diffusivity coefficient is the same for mass and momentum transport, and $U_x'(y)= dU_x/dy$. 

By inserting Eq.~(\ref{eq:Reynolds}) into Eq.~(\ref{eq:c2}), we arrive at an equation for the time-averaged concentration, which can be readily integrated to a formal solution given as  
\begin{eqnarray} \label{eq:C}
\overline{C} =\int_0^y\frac{dy'}{\epsilon(y')}\left(\int_0^{y'} dy'' U_x(y'')\right)\frac{\partial \overline{C}}{\partial \xi} +\frac{\partial\overline{C}}{\partial \xi}\xi,
\end{eqnarray}
under the assumption that $\partial \overline{C}/\partial \xi = \textrm{constant}$. This means that mean concentration field has reached a steady state profile (with a linear decrease from the source) sufficiently far away from the pipe's inlet, where it was injected.

The longitudinal diffusivity can be calculated from the advective flux averaged over the width of the pipe, 
\begin{equation}
Q_L = -K_L\partial \overline{C}/\partial \xi=H^{-1}\int_0^H dy U_x(y)\overline{C}(y,\xi)\end{equation}
or, equivalently, by using the mean concentration $\overline{C}(y,\xi)$ from Eq.~(\ref{eq:C}), as
\begin{eqnarray}\label{eq:KLadv}
K_L = -\frac{1}{H}\int_0^HdyU_x(y)\int_0^y \frac{dy'}{\epsilon(y')}\int_0^{y'} dy'' U_x(y'').
\end{eqnarray}
Turbulent fluctuations also contribute to a diffusive flux and the associated turbulent diffusivity is the average across the width of the channel of the local turbulent diffusivity, i.e. $K^{turb} = H^{-1}\int_0^Hdy\epsilon(y)$. However, it is known that longitudinal dispersion by mean flow advection overcasts the turbulent and the molecular dispersions~\cite{fischer1973longitudinal}, and it will be this main contribution that we attribute to the longitudinal dispersion. Nonetheless, as seen from Eq.~(\ref{eq:KLadv}), turbulence affects $K_L$ through its nontrivial dependence on the mean velocity profile $U_x(y)$ and the turbulent shear stress $\tau(y)$.     

The other important ingredient in Taylor's theory is the assumption that the mean velocity $U_x(y)$ and the turbulent shear stress $\tau(y)$ can be expressed in terms of their universal profiles in the outer layer. That is true in the asymptotic limit of $Re\rightarrow\infty$, and it means that, when measured in typical units related to the wall friction velocity $U_w$ and the width of the channel, $H$, these functions can be expressed as~\cite{Taylor1954} 
\begin{eqnarray}
U_x(y) &=& U_\infty - U_w \hat U(\hat y)\\
\tau(y) &=& \rho U_w^2 \hat\tau(\hat y)
\end{eqnarray}
where $\hat y= y/H$, $U_\infty$ is a reference velocity in the bulk, $\rho$ is the fluid density, and $\hat U(\hat y)$ and $\hat \tau(\hat y)$ are universal, dimensionless functions. As a consequence, the corresponding change of variables in the integrals from Eq.~(\ref{eq:KLadv}) implies that the longitudinal diffusivity is also measured in units of the rescaling variables, $K_L = \alpha_\infty U_wH$, up to a constant prefactor $\alpha_\infty$ given by 
\begin{equation}\label{eq:alpha}
\alpha_\infty =-\int_0^1d\hat y \Delta U_x(\hat y)\int_0^{\hat y} d\hat y'\frac{\hat U'(\hat y')}{\hat\epsilon(\hat y')}\int_0^{\hat y'} d\hat y'' \Delta U_x(\hat y''),
\end{equation}
where $\Delta U_x(\hat y) = U_\infty/U_w-\hat U_x(\hat y)$. The numerical value of $\alpha_\infty$ thus depends on the actual shape of the universal profiles for channel flow. From the dependence of the friction factor with $Re$, i.e. $f\sim Re^{-\beta}$, we can then predict that, in this asymptotic regime of $Re\rightarrow\infty$, the diffusivity should scales as $K_L\sim Re^{-\beta/2}$. In the momentum transport theory~\cite{gioia2006turbulent}, also discussed in the next section, the scaling exponent $\beta$ of the friction factor is related to the Kolmorogov scaling exponent of the turbulent energy spectrum. 

Taylor's theory of dispersion neglects the contribution from the wall region and becomes valid for $Re>2\times 10^4$, at least for pipe flows~\cite{tichacek1957axial,flint1969longitudinal,ekambara2003axial}. In the region of the inner boundary layers, the typical units change to the wall variables, frictional velocity $U_w$ for the mean velocity, and the viscous lengthscale $l_w = \nu/U_w = H/(Re\sqrt{f})$ for the distance to the wall $y$ using the friction factor $f=U_w^2/U_0^2$. Then, the universal velocity and shear stress profiles written in the wall variables read as
\begin{eqnarray}
U_x(y) &=& U_w \tilde U(\tilde y),\\
\tau(y) &=& \rho U_w^2 \tilde\tau(\tilde y)
\end{eqnarray}
with $\tilde y = y/l_w= y Re\sqrt{f}/H$. The inner boundary layer extends up to $\tilde y_c=y_c Re\sqrt{f}/H$ above which it crosses over to the outer boundary layer scaling. The important point here is that because the inner and outer boundary layers are represented by different typical lengthscale, the uniform change of variable in Eq.~(\ref{eq:KLadv}) for the integration domain is now replaced with $y\rightarrow y/l_w$ for $y<y_c$ and $y\rightarrow y/H$ for $y>y_c$, where the thickness of the wall region $y_c$ is taken as a scale parameter. Hence, the longitudinal diffusivity measured in units of $U_0H$ is given by 
\begin{eqnarray}\label{eq:KLscaling}
&&\frac{K_L}{HU_0} = \sqrt{f}\alpha\left(\frac{\tilde y_c}{Re\sqrt{f}}\right)\nonumber\\
&+&\frac{1}{Re}g_1\left(\frac{\tilde y_c}{Re\sqrt{f}}, \tilde y_c\right)+\frac{1}{Re^2 \sqrt{f}}g_2\left(\tilde y_c\right), 
\end{eqnarray}
where the scaling functions $\alpha$, $g_1$, and $g_2$ are given below. 
\begin{eqnarray}\label{eq:scalingfunc1}
\alpha\left(\lambda\right) = -\int_{\lambda}^1d\hat y\Delta\hat U(\hat y)\int_{\lambda}^{\hat y}d\hat y' \frac{\hat U'(\hat y')}{\hat\tau(\hat y')}\int_{\lambda}^{\hat y'}d\hat y''\Delta\hat U(\hat y''),\nonumber\\ 
\end{eqnarray}
where $\lambda= y_c/H=\tilde y_c/(Re\sqrt{f})$. We see that in the limit of high $Re$-numbers, $\lambda\rightarrow 0$ and $\alpha(\lambda)\rightarrow\alpha_\infty$. 
\begin{eqnarray}\label{eq:scalingfunc2}
&&g_1\left(\lambda,\tilde y_c\right) = \int_{\lambda}^1d\hat y\Delta\hat U(\hat y)\int_{0}^{\tilde y_c}d\tilde y' \frac{\tilde U'(\tilde y')}{\tilde\tau(\tilde y')}\int_{0}^{\tilde y''}d\tilde y''\tilde U(\tilde y'')\nonumber\\
&-&\int_{\lambda}^1d\hat y\Delta\hat U(\hat y)\int_{\lambda}^{\hat y}d\hat y' \frac{\hat U'(\hat y')}{\hat\tau(\hat y')}\int_{0}^{\hat y_c}d\tilde y''\tilde U(\tilde y''), 
\end{eqnarray}
and 
\begin{eqnarray}\label{eq:scalingfunc3}
g_2\left(\tilde y_c\right) = \int_{0}^{\tilde y_c}d\tilde y\tilde U(\tilde y)\int_{0}^{\tilde y_c}d\tilde y' \frac{\tilde U'(\tilde y')}{\tilde\tau(\tilde y')}\int_{0}^{\tilde y''}d\tilde y''\tilde U(\tilde y'').\nonumber\\ 
\end{eqnarray}
From Eq.~(\ref{eq:KLscaling}), we notice that the first term is always going to dominate for large $Re$-numbers, and we recover Taylor's original asymptotic scaling $K_L\sim Re^{-\beta/2}$. However, at intermediate $Re$-numbers the other scaling behavior due to boundary layer effects come into play and we may see a cross-over to a different scaling regime as the $Re$ number is lowered. 
 
Owing to the direct connection between friction factor and longitudinal diffusivity, we expect different asymptotic scaling laws of $K_L$ with $Re$ number, corresponding to different turbulent cascades.

\section{Connection to the energy spectrum} 
The momentum transfer model for friction factor proposed in Ref.~\cite{gioia2006turbulent} links the asymptotic scaling laws of the friction factor with $Re$ and wall roughness with the turbulent energy spectrum. The idea is that the momentum transfer from the bulk to the wall is mostly enabled by eddies of sizes comparable to a typical lengthscale $s$, determined by the Kolmogorov lengthscale and the typical size of the wall roughness. To extract the scaling with $Re$ number, the limit of zero wall roughness is taken where $s$ is determined by the Kolmogorov scale. From a sea of turbulent eddies, those that are straddled near the wall and of size$~s$ are contributing most to the wall shear stress $\tau_w$, hence $\tau_w\sim \rho U_0u_s$, where $u_s$ is the typical swirling velocity of an eddy of size $s$ and estimated by integrating the kinetic energy up to that scale or equivalently 
\begin{equation}
u_s\sim\left(\int_{s^{-1}}^\infty dk E(k)\right)^{1/2}.
\end{equation}
Since the friction factor is a dimensionless form of the wall shear stress, $f = \tau_w/\rho U_0^2$, then 
\begin{equation}
f\sim U_0^{-1} \left(\int_{s^{-1}}^\infty dk E(k)\right)^{1/2}, 
\end{equation}
and its scaling with $Re$ number emerges from $s(Re)$. In the inverse energy cascade we have that $E(k)\sim k^{-5/3}$ and the Kolmogorov lengthscale $s\sim Re^{-3/4}$, which implies that $f\sim Re^{-1/4}$. However, in the enstrophy cascade regime where $E(k)\sim k^{-3}$ and $s\sim Re^{-1/2}$, the model predicts that the friction factor scales instead as $f\sim Re^{-1/2}$, e.g.~\cite{guttenberg2009friction}. These scaling laws have been measured both numerically~\cite{gioia2006turbulent} and in soap film experiments~\cite{tran2010macroscopic}. The relationship between $K_L$ and $f$ from Eq.~(\ref{eq:KLscaling}) implies that the asymptotic ($Re\rightarrow\infty$)  scaling behavior of $K_L(Re)$ with $Re$ is connected with the turbulent energy spectrum, i.e. $K_L\sim Re^{-1/8}$ in the energy cascade regime and $K_L\sim Re^{-1/4}$ in the enstrophy cascade regime.

\begin{figure}[t]
\begin{centering} 
 \includegraphics[width=0.8\columnwidth]{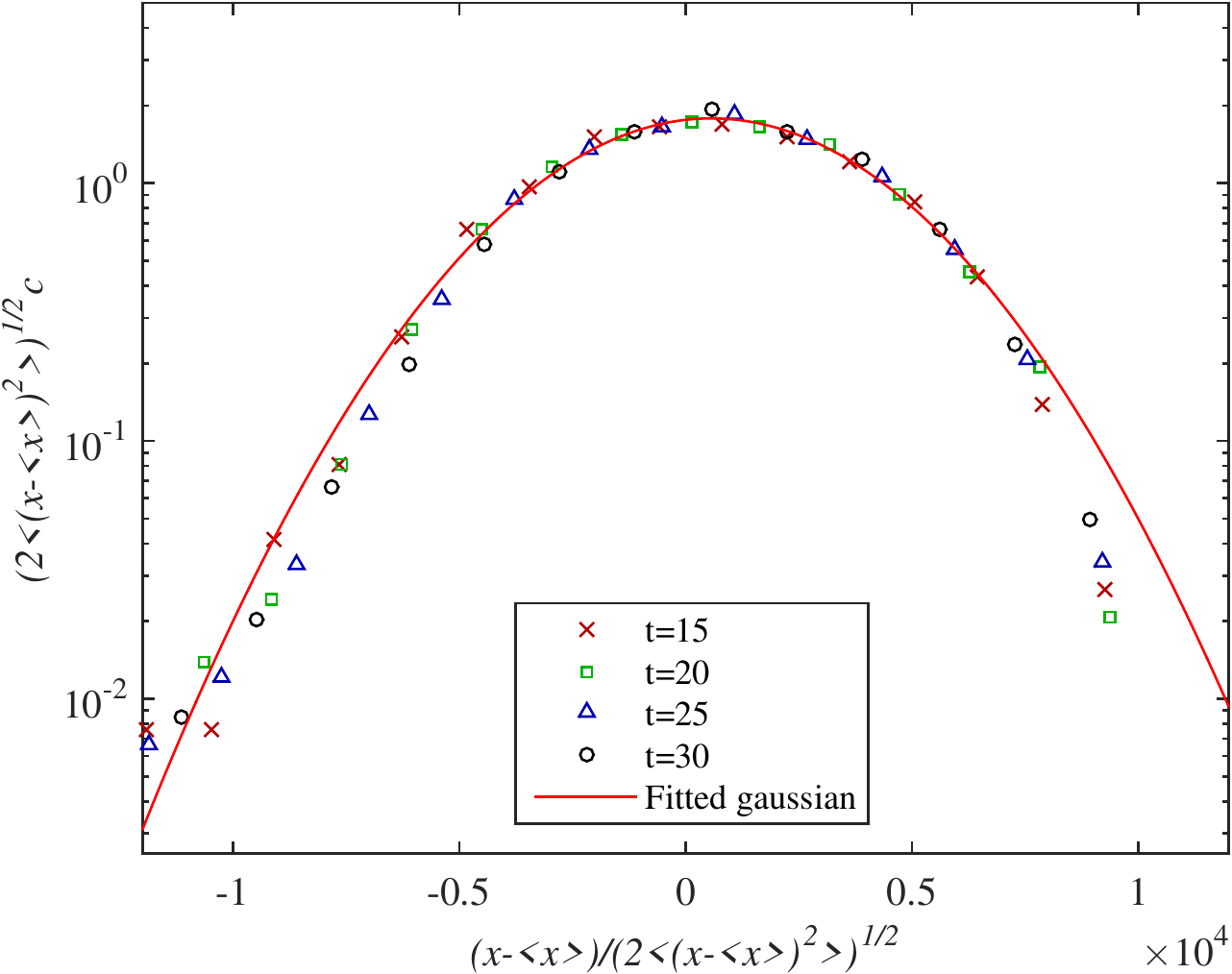}
 \caption{\label{fig:4} (Color online) Longitudinal concentration of tracers in the comoving frame. Gaussian fit of the core distribution.}
\end{centering}
\end{figure}


\section{Lattice Boltzmann simulations:} To check these scaling laws is rather challenging for several computational and theoretical reasons. Numerically, it is difficult to simulate statistically stationary turbulent flows at very high Re numbers due to the drag on the wall. Theoretically, it is not fully understood the mechanism of generating single or coexisting inertial cascades in wall-bounded turbulence~\cite{jimenez2011cascades}. Nonetheless, soap film experiments~\cite{rutgers1998forced,bruneau2005experiments} accompanied by few numerical simulations~\cite{bruneau2005experiments} show evidence that 2D turbulence can be excited by wall roughness such that an inverse energy cascade is coexisting with a forward enstrophy cascade. This is different from the grid generated turbulence bounded by smooth walls, where a single cascade of enstrophy is developed~\cite{kellay2002two}. A turbulent spectrum with a single inverse energy cascade has been measured in experiments where the soap film is pierced at the inlet with a cylindrical rod and flows between two wires, one of which is made rough~\cite{kellay2012testing}. 

\begin{figure}[t]
\begin{centering} 
 \includegraphics[width=0.8\columnwidth]{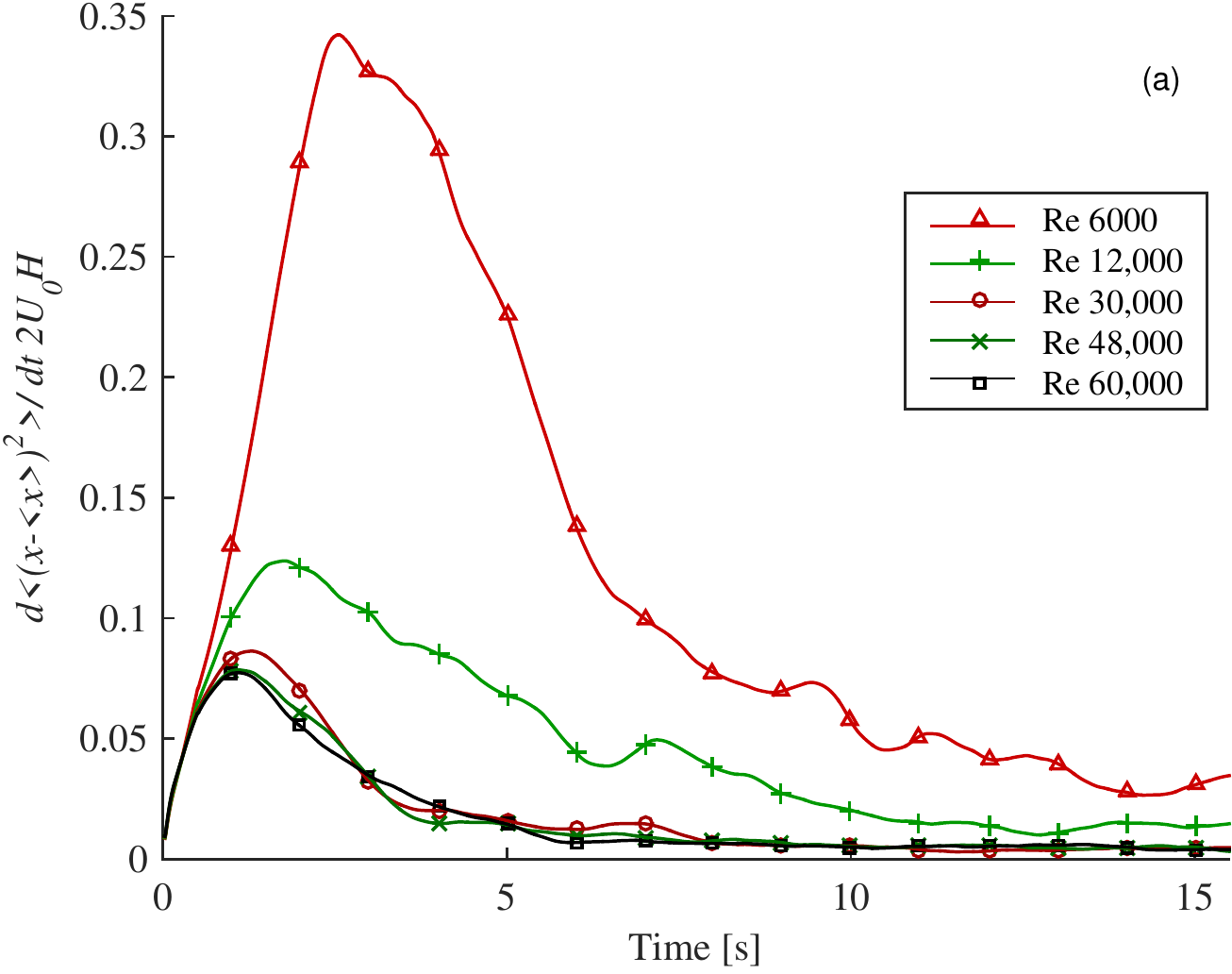}
\includegraphics[width=0.8\columnwidth]{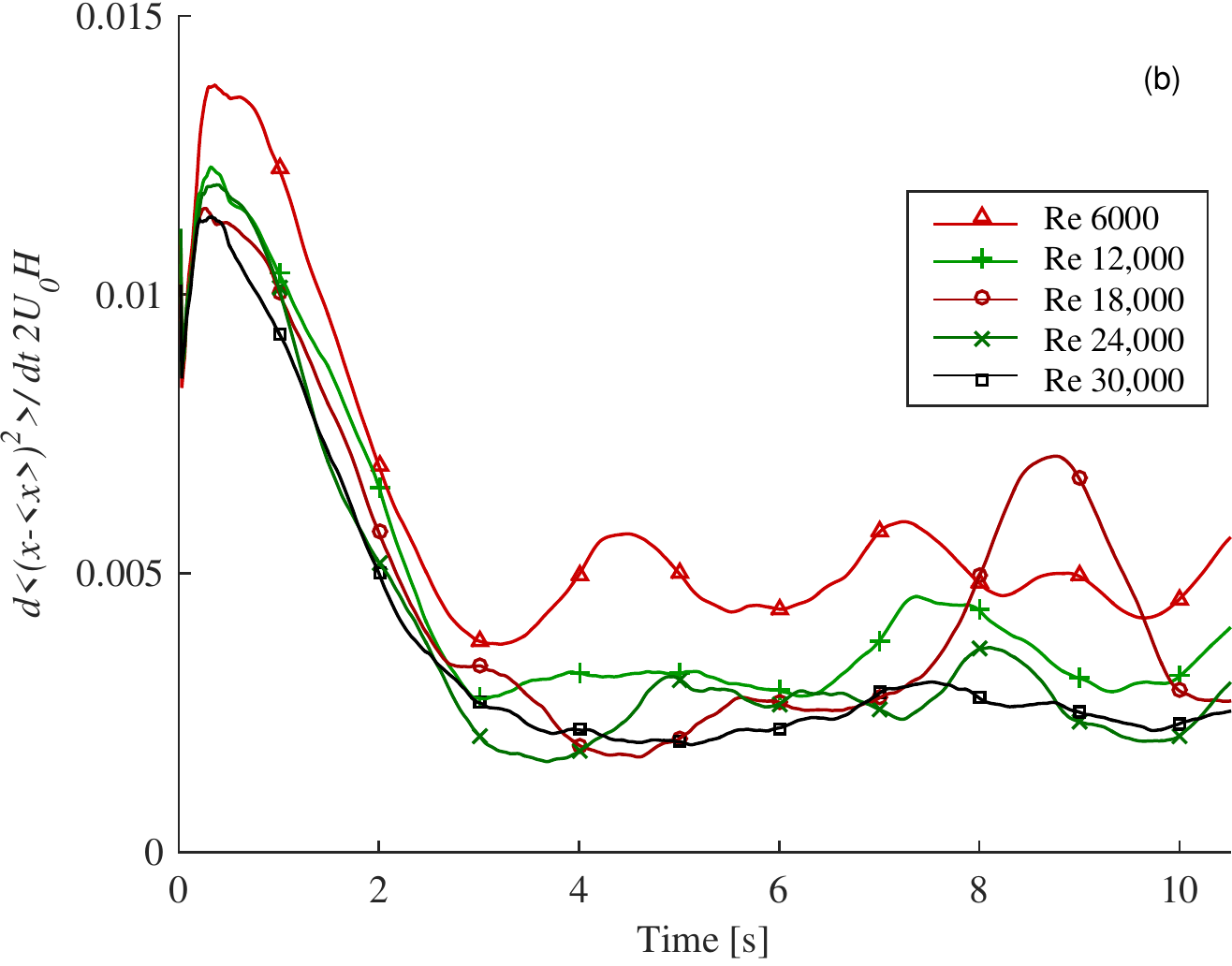}
 \caption{\label{fig:3} (Color online) Time-dependence of the rescaled longitudinal diffusivity measured from the mean-squared displacement of passive tracers at different $Re$ numbers measured in a turbulent flow generated (a) from wall roughness, and (b) behind a grid.}
\end{centering}
\end{figure}

We use direct numerical simulations of a turbulent channel flow using the two-dimensional incompressible formulation of the lattice Boltzmann model type LBGK (D2Q9)~\cite{succi01}. Periodic conditions are applied at the inlet and outlet, and no-slip walls on the long sides of the channel are implemented via the bounce-back rule~\cite{succi01}. Numerical stability and the flow incompressibility depend on the grid resolution that is $Re$-number dependent as discussed in e.g. Ref.~\cite{he1997lattice}. More details on the numerical lattice Boltzmann model can be found in Refs.~\cite{hawkins2013precipitation,hawkins2014hydrodynamic}. For the turbulent flow induced by wall roughness, $5$ semicircular asperities of equal size are randomly distributed along the top and bottom walls of the pipe. In the case of the grid turbulence, $5$ circular asperities with size $r/H=0.04$ are uniformly spaced across the pipe at a given distance from the inlet (movies in the Supplementary Material). As an initial condition, we start with a laminar flow profile. During the time evolution, the laminar velocity field gets perturbed either by the wall asperities or the transverse grid, until turbulent fluctuations take over. All the statistical analysis is done after this transitory time. The no-slip condition generates a wall shear stress or drag which cause a gradual dissipation of fluid flow. We however measured that the total kinetic energy decays with time as $1/t$, and in this case the scaling properties of transport and dispersion in a steady-state flow should also remain valid for the decaying turbulence when the time-dependence is scaled away~\cite{huang1995velocity}.  

We calculate the statistics of single-particle dispersion using passive tracers advected with the local fluid velocity, $\dot{\mathbf{x}}^{(i)} = \mathbf{u}(\mathbf{x}^{(i)},t)$,  for $i=1,\cdots,N$ where we have $N= 10^4$ total number of particles. The local velocities at the particles' locations are determined using a second-order interpolation of the lattice velocity field, and the Lagrangian advection is performed by the forward Euler's scheme.

{\section{Discussion and conclusions} We compute the mean square displacement from the particles' positions. Alternatively, it can also be estimated from the spread of the number of particles at a given location along the channel. This is shown in Figure~~\ref{fig:4}, there the concentration of particles averaged over the width of the channel is plotted at different times as function of the position along the channel. We notice that the longitudinal concentration can be approximated by a Gaussian distribution with the variance given by the mean square displacement. However, there are deviations in the tail distribution that maybe related to a decaying turbulence and a time dependent mean velocity $U_0(t)$. 

Time-dependent longitudinal diffusivity is computed as the time-derivative of the mean square longitudinal displacement, $\frac{d}{2dt}\langle (x-\langle x\rangle)^2\rangle$. To emininate the effect of the decaying turbulence, we rescale it by typical units $U_0(t)H$. Figure~\ref{fig:3} shows the temporal dependence of this rescaled diffusivity for different $Re$ numbers in the \emph{roughness-induced} turbulent (panel (a)) and in the \emph{grid-generated} turbulence (panel (b)). In the long-time limit, it fluctuates about a constant value given by the dimensionless diffusion coefficient 
\begin{equation}
\frac{K_L}{U_0 H} = \frac{1}{2U_0H}\lim_{t\rightarrow\infty}\frac{d}{dt}\langle (x-\langle x\rangle)^2\rangle.
\end{equation} 
Turbulent cascades in 2D turbulence can be inferred from the scaling behavior of the energy spectrum across the inertial scales. With lattice Boltzmann simulations, we are able to compute the Eulerian $\Phi_E(\omega)$ and Lagrangian $\Phi_L(\omega)$ frequency spectra (presented in Figure~\ref{fig:2}) from the temporal signal of the transverse velocity with zero mean, which gives us a proxy of turbulent fluctuations without the effect of the mean flow. The Lagrangian frequency spectrum $\Phi_L(\omega)$ is given by the power spectrum of the transverse velocity along particles' trajectories, whereas the Eulerian frequency spectrum $\Phi_E(\omega)$ is calculated as the power spectrum of the temporal velocity signal at a fixed measurement point in space. We find that, at sufficiently high Re numbers, different scaling behaviors of the frequency spectra emerge corresponding to different turbulent cascades developed in the roughness-generated turbulence and the grid-generated turbulence. This is also shown in Figure~(\ref{fig:2}).

The scaling regime $\Phi_L(\omega)\sim \omega^{-2}$ is consistent with the $E(k)\sim k^{-5/3}$ law for an \emph{inverse energy cascade}~\cite{tennekes1972first}. On dimensional analysis ground and based on the statistical independence of the small-scale turbulence from the large-scale structures, this follows from the relation of the eddy wavenumber $k$ with its typical turnover frequency $\omega\sim \epsilon^{1/3}k^{2/3}$, where $\epsilon$ is the constant energy dissipation rate, and by expressing the kinetic energy contained in an eddy in equivalent ways $kE(k)=\omega \Phi_L(\omega)$. We find that this scaling is dominant in the \emph{roughness-induced turbulence} for large $Re$ numbers as seen in Fig.~\ref{fig:2} (panel (a)). At large $\omega$'s, there is a cross-over to a power spectrum steeper that $-2$ suggesting a coexisting enstrophy cascade~\cite{rutgers1998forced}. For the Eulerian spectra $\Phi_E(\omega)\sim \omega^{-5/3}$ is consistent with the $-5/3$'s law using the~\lq random sweeping\rq~hypothesis of the small-scale eddies by the large-scale eddies~\cite{tennekes1975eulerian}.   

\begin{figure}[t]
\begin{centering} 
 \includegraphics[width=\columnwidth]{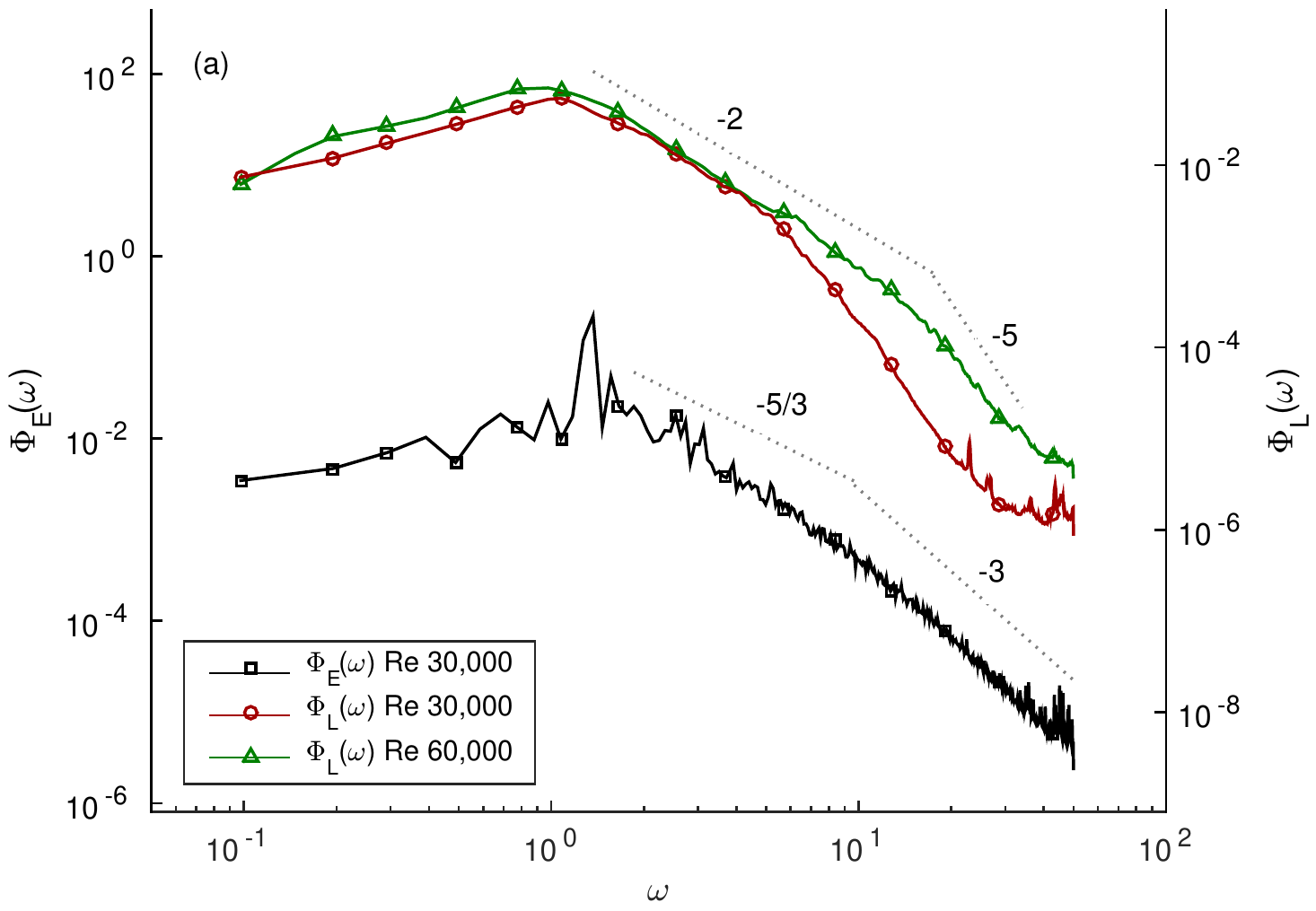}
 \includegraphics[width=\columnwidth]{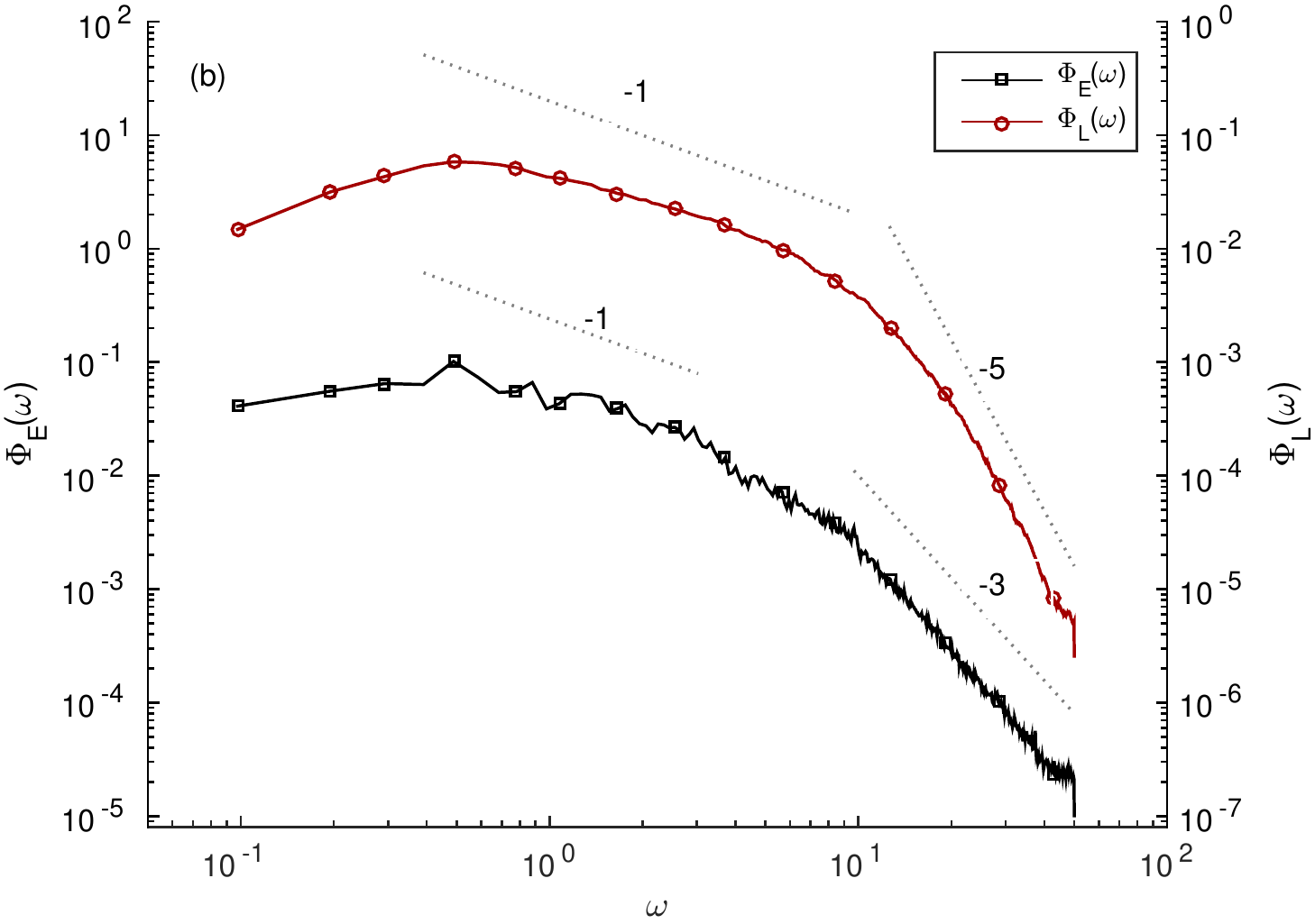}
 \caption{\label{fig:2} (Color online) Lagrangian frequency spectrum $\Phi_L(\omega)$ and Eulerian frequency spectrum $\Phi_E(\omega)$ of the transverse velocity fluctuations in a turbulent flow at $Re= 30000$ that develops (a) from the wall roughness, and (b) behind a grid.
}
\end{centering}
\end{figure}

For the enstrophy cascade, we lack a simple dimensional prediction since the turnover frequency depends solely on the enstrophy dissipation rate $\eta$ as $\omega\sim \eta^{1/3}$, hence an invariant across the inertial eddies. However, based on previous studies of 2D turbulence~\cite{rutgers1998forced,guttenberg2009friction}, we expect a \emph{direct enstrophy cascade} dominating the energy spectrum in the \emph{turbulence developed behind a grid}. $\Phi_L$ at high $Re$ numbers, as seen in panel (b) of Fig.~\ref{fig:2}, scales with an exponent steeper that $-2$, approaching $\Phi_L(\omega)\sim \omega^{-5}$ at large $\omega$'s, consistent with other studies of Lagrangian statistics in homogeneous 2D turbulence~\cite{babiano1987single,provenzale1995single}. $\Phi_E(\omega)$ scales similarly to the wavenumber spectrum $E(k)$ with a $\omega^{-3}$-scaling at large $\omega$'s corresponding to an enstrophy cascade. We invoke the~\lq random sweeping\rq~hypothesis for this scaling similarity, although this is not fully understood. We notice that both frequency spectra in the grid turbulence develop a $\omega^{-1}$ scaling at lower frequencies, where to cascade develops and turbulent fluctuations are represented by well-separated vortices that are moving in the velocity field induced by each other, without merging or splitting~\cite{novikov1975dynamics}.     

\begin{figure}[t]
\begin{centering} 
 \includegraphics[width=\columnwidth]{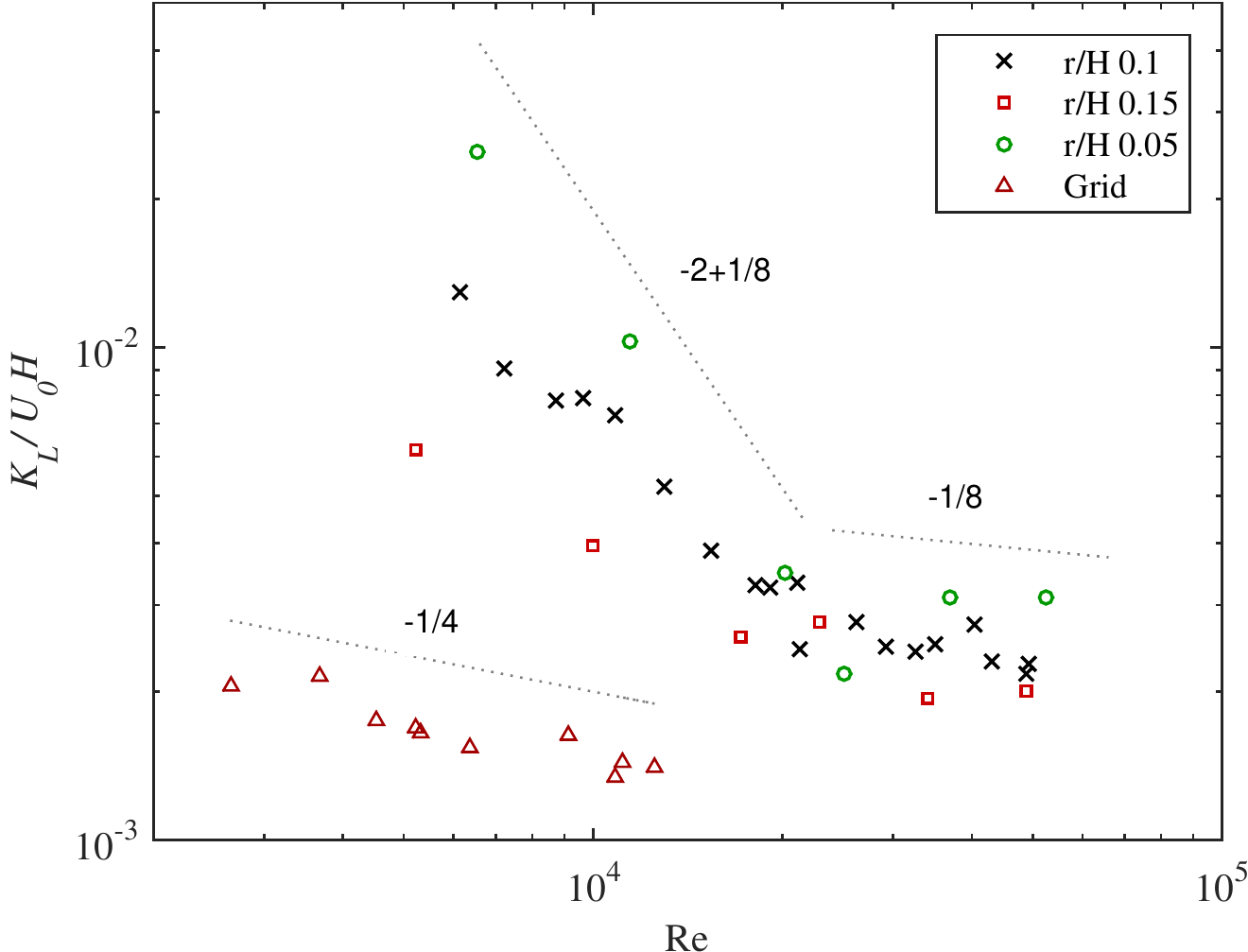}
 \caption{\label{fig:5} (Color online) Longitudinal diffusivity $K_L$ with $Re$ numbers for the grid and rough-wall induced turbulence.}
\end{centering}
\end{figure}

The constant diffusivity $K_L/(U_0 H)$, measured in the typical units $U_0H$, varies with the $Re$ number in a fundamentally different manner depending on the dominating turbulent energy spectrum. In Figure~\ref{fig:5}, we present the scaling laws of the longitudinal diffusivity consistent with our predictions, albeit the range is very restricted. The longitudinal diffusivity in a turbulent flow with rough walls is computed as a function of $Re$ for three different sizes of wall roughness, i.e. $r/H=0.05,0.15,0.1$. Even though we are computationally limited to explore very large $Re$ numbers, we observe that for $Re>10^4$, the asymptotic scaling law $K_L/(U_0 H)\sim Re^{-1/8}$ predicted from an \emph{inverse energy cascade} becomes apparent. Admittedly, the presence of this scaling law with a small exponent and on a narrow range is debatable, and needs to be explored more both experimentally and numerically. At intermediate $Re<10^4$, a different scaling regime is observed consisted with our predictions in Eq.~(\ref{eq:KLscaling}) when the boundary layers are included. For the grid-generated turbulence dominated by an \emph{enstrophy cascade}, the turbulent fluctuations are stronger and the boundary layer effect is not as evident. In fact, we see that the asymptotic scaling law with $Re$ number, $K_L/(U_0 H)\sim Re^{-1/4}$, is already present for $Re$ below $10^4$ as long as the turbulent  flow is developed.  

To conclude, we have shown that in the Reynolds analogy between mass and momentum transfer, a spectral link is manifested for scalar transport properties. This dependence on the turbulent cascades determines the asymptotic scaling law of $K_L(Re)$, and remains to be validated in future pipe and channel flow experiments.

{\emph{Acknowledgments:} We are grateful to Nigel Goldenfeld for insightful discussions. This study was supported by a doctoral fellowship from MINSC, a European Marie Curie Initial Training Network.

\bibliographystyle{apsrev4-1}
\bibliography{refs}

\end{document}